\documentclass[prl,twocolumn, showpacs,groupeaddress]{revtex4}
\usepackage{amsmath}
\usepackage{amsfonts}
\usepackage{graphicx}
\usepackage{times}

\begin{document}

\title{Geometric phase distributions for open quantum systems}
\author{K.-P. Marzlin}
\author{S. Ghose}
\author{B.C. Sanders}
\affiliation{
\mbox{Institute for Quantum Information Science, University of Calgary,~2500 University Drive NW, Calgary, Alberta T2N 1N4, Canada}}

\begin{abstract}
In an open system, the geometric phase should be described by a
distribution. We show that a geometric phase distribution for open system
dynamics is in general ambiguous, but the imposition of reasonable
physical constraints on the environment and its coupling with the system
yields a unique geometric phase distribution that applies even for mixed
states, non-unitary dynamics, and non-cyclic evolutions.
\end{abstract}


\date{\today}

\maketitle

{\em Introduction.}--- 
The geometric phase (GP) \cite{berry84,pancharatnam56}
identifies the portion of an overall (abelian  or 
non-abelian \cite{anandan88b}) phase shift of a quantum
state that is due to the path of the state through projective Hilbert
space; the GP and
the dynamic phase combine to give the aggregate phase shift of the state
that may be inferred by interferometric or other phase-sensitive methods.
GP theory has been rigorously formulated for the general case of
non-adiabatic \cite{aharonov87}, non-cyclic \cite{samuel88}, and
non-unitary evolution (without quantum jumps)
\cite{uhlmann86,pati95a,pati95b} 
of a pure state, but the importance of GP in realistic
systems, for example in the context of adiabatic quantum computation
\cite{adiqc,pachos99}, has
motivated recent important research into GP in open systems
\cite{gamliel89,ellinas89,sjoquist00,fonseca02,peixoto03,singh03,fu04,whitney04}.
Quantum jump (or trajectory) analyses have been applied to 
certain physical systems, which
show how the GP for a closed system can be modified under open system
dynamics \cite{fuentes03,fuentes03b}, 
and a rigorous Kraus operator approach has been applied to
define GP for general open system evolution 
\cite{kult03}. These
studies note the importance of GP beyond closed-system, unitary evolution
of pure states. 

In this paper we argue that a complete description of 
abelian GP in open systems has to identify
the appropriate measure of phase distribution. We develop a
theory of GP distributions for mixed states, non-unitary dynamics, 
and non-cyclic evolutions. We show that,
without further contraints, the
GP distribution is ambiguous: an operational definition of GP that would
resolve this ambiguity is not attainable because the GP is a non-linear
functional of the state.
The ambiguity of the phase distribution is rather subtle: we
show that previous definitions of GP distributions and its spread 
implicitly assume a particular form of phase distribution. 
The imposition of reasonable
physical contraints on the environment and its coupling with the system
yields a unique GP distribution by taking the decomposition
of the density matrix \cite{kult03} into account.

{\em Definition of geometric phase distributions for an open system.}---
Interferometric or other phase-sensitive measurements allow
inference of the phase shift of a state, but separating geometric and
dynamic components of the phase is not straightforward.
For a pure state $|\psi(t)\rangle = V(t) |\psi (0)\rangle \in {\cal H}$ 
which is propagated by  an arbitrary time-dependent
evolution operator $V(t)$  (not necessarily cyclic or unitary),
the mathematical definition of the geometric phase functional
$\beta [\psi]$ is given by \cite{samuel88}
\begin{equation} 
  \text{e}^{\text{i}\beta [\psi]} 
  \equiv \frac{\text{Z}[\psi ]}{|\text{Z}[\psi] |}\quad , 
  \quad \text{Z}[\psi] \equiv  
  D[\psi]\; \langle \psi(0) | \psi(t) \rangle  \; ,
\label{samuelGP}
\end{equation}
which is meaningful only for $Z[\psi] \neq 0$.
The dynamic phase functional of $|\psi \rangle$
\begin{equation} 
  D[\psi] \equiv \exp \left ( -\text{i} \int_0^t \text{d}t^\prime \mbox{Im} 
  \frac{\langle \psi(t^\prime ) | \dot{\psi}(t^\prime ) \rangle 
                }{\langle \psi(t^\prime ) | \psi(t^\prime ) \rangle}
 \right ) 
\label{ddef}\end{equation} 
removes the dynamic phase from the total phase shift
associated with $V(t)$.
In some cases, the dynamic phase can be eliminated via interferometry of a
state that follows a superposition of two paths, with the
dynamic phase along each path being the additive inverse of the other
\cite{webb99,sanders01}, but this cancellation of
dynamic phase is not always achievable.

Non-unitary evolution of the type $V(t)$ 
may not satisfy the axioms of completely positive (CP) maps,
which guarantee that a positive-definite operator on Hilbert space such as
the density operator~$\rho$ is mapped to a positive definite operator with
identical trace, and linearity is preserved. 
Thus the GP should be established for general CP maps, not
just for non-unitary evolution \cite{ericsson03,pati03,peixoto03}. 
A physical picture for the CP emerges by
considering a system S with Hilbert space $\mathcal{H}_\text{S}$ and a
reservoir R (or set of ancillae) with Hilbert space $\mathcal{H}_\text{R}$,
and joint Hilbert space
$\mathcal{H}=\mathcal{H}_\text{S}\otimes\mathcal{H}_\text{R}$.
At some initial time $t=0$ the joint density factorizes,
$\rho_\text{SR}(0)=\rho_\text{S}(0)\otimes\rho_\text{R}(0)$, 
and the unitary operator
$U_\text{SR}(t)$ of the system+reservoir (S+R) is generated by a hamiltonian
$H_\text{SR}(t)$ such that 
$\rho_\text{SR}(t)=U_\text{SR}(t)\rho_\text{SR}(0)U_\text{SR}^\dagger(t)$.
Dynamics for S alone is obtained by tracing over R; i.e.
\begin{equation}
\rho_\text{S}(t) 
        = \text{Tr}_\text{R}\left(U_\text{SR}
        \rho_\text{SR}(0)U^\dagger_\text{SR}\right)
\end{equation}
is a CP mapping, which  can be decomposed into a sum of 
mappings corresponding to various measurement records obtained by readouts
of the environments. 
For a particular initial state
$\rho_\text{S}(0) = \sum_s q_s |\psi_s \rangle \langle \psi_s |$ 
and 
$\rho_\text{R}(0) = \sum_r p_r |r \rangle \langle r |$ 
this Kraus decomposition corresponds to
\begin{eqnarray}
   \rho_\text{S}(t) &=& 
   \sum_{b_\text{R},r,s} p_r q_s  U_{r,b_\text{R}}(t)
   | \psi_s \rangle \langle \psi_s | 
   U_{b_\text{R},r}^\dagger (t)
\label{rdmdef} \end{eqnarray} 
with Kraus operators 
$ U_{b_\text{R}, r} \equiv \langle b_\text{R} | U_\text{SR}(t) | r\rangle $, 
where
$\{ |b_\text{R} \rangle \} $ is some basis for the Hilbert space 
of the reservoir.
This corresponds to an incoherent mixture of non-unitarily evolving states 
$  |\psi_{b_\text{R}, r,s}\rangle \equiv U_{b_\text{R},r}(t) |\psi_s \rangle$,
weighted with the initial probabilities $p_r q_s$. 
For each individual state the GP is given by Eq.~(\ref{samuelGP})
with $\text{Z}[\psi_{b_\text{R}, r,s}] \propto 
\langle\psi_{b_\text{R}, r, s} (0) |\psi_{b_\text{R} ,r, s}(t)\rangle$.
Because $| \psi_{b_\text{R}, r, s} (0) \rangle = 
\langle b_\text{R} | r \rangle |\psi_s \rangle $, it is obvious that
$\text{Z}[\psi_{b_\text{R}, r,s}]=0$ whenever the basis state 
$| b_\text{R} \rangle $ is orthogonal to the reservoir state $|r \rangle $,
so that the associated GP is not well defined. To avoid this problem
we start by choosing a different basis set 
$\{ |b_\text{R} (r)  \rangle  \} $ for
each term in the sum over $r$ in Eq.~(\ref{rdmdef}), such that 
$\langle b_\text{R} (r) | r \rangle = \delta_{b_\text{R},0}$.
Consequently, when one introduces a distribution for complex numbers
of the form
\begin{equation} 
  P(z) = \sum_{b_\text{R},r,s} w(r,s,b_\text{R})
   \delta (z - \text{Z}[\psi_{b_\text{R}, r,s}]) \; , \quad z \in \mathbb{C} \; ,
\end{equation} 
with $w(r,s,b_\text{R})$ being arbitrary weight functions, 
all terms with $b_\text{R} \neq 0$
do not contribute to any moment 
$\langle z^n \rangle = \int  z^n P(z) \text{d}^2z$ 
and can therefore be omitted. Hence, it is
sufficient to keep only terms with $b_\text{R} = 0$ 
so that the complex number distribution induced by Eq.~(\ref{rdmdef})
becomes
\begin{equation} 
  P_\text{Z}(z) \equiv \sum_{r,s} p_r q_s  \delta (z - \text{Z}[\psi_{r,s}]) ,
\label{pz}\end{equation} 
with $ | \psi_{r,s}(t)\rangle =  
\langle r |U_\text{SR}(t)| r \rangle \; |\psi_s\rangle $.
For a total system S+R with a continuous spectrum the sums 
in this distribution would be replaced by integrals.

The definition for a corresponding GP distribution faces the same subtleties
that arise for any phase distribution. Usually, a 
positive operator-valued measure (POVM) is introduced to
describe a phase distribution for quantum systems. However, 
through the dynamic phase functional $ D[\psi] $, 
the GP depends non-linearly on the states of the system,
so that it is generally not possible to provide a GP POVM based on 
linear operators. 
Instead, one has to construct a GP phase distribution differently. 
We investigate two natural definitions of GP distributions.
The first possibility to introduce a GP distribution
is to derive the GP directly from the complex number distribution 
$P_\text{Z}$ of Eq.~(\ref{pz}); for example the
mean GP is given by the first moment 
\cite{betaMeanRemark}
\begin{equation} 
  \text{e}^{\text{i} \langle \beta \rangle  } \equiv 
  \frac{\langle z \rangle_\text{Z}}{|\langle z \rangle_\text{Z}|} = 
  \frac{\sum_{r,s } p_r  q_s Z[\psi_{ r, s}]
  }{ | \sum_{r,s } p_r  q_s Z[\psi_{ r, s}] | } \; .
\label{berrydef}
\end{equation}
By definition, 
$ D[\psi_{r,s}] $  becomes unity if each 
$|\psi_{r,s} \rangle $ is parallel transported. 
In this case definition 
(\ref{berrydef}) can be rewritten as 
$ \exp (\text{i} \langle \beta \rangle ) = 
\mbox{Tr} ( U_\text{SR}(t) \rho_\text{SR}(0))/
| \mbox{Tr} ( U_\text{SR}(t) \rho_\text{SR}(0))| $,
which coincides with the definition given in Ref.~\cite{sjoquist00}.

A second possibility to introduce a GP distribution is 
motivated by Holevo's approach to moments of a phase 
distribution \cite{holevo84}.
While the definition (\ref{berrydef}) depends on the modulus of
$Z[\psi]$, too, this is not necessarily desirable. Instead, one can
introduce phase distribution of the form
\begin{equation} 
  P_\text{H}(s) = \sum_{r,s} p_r q_s  \;
   \delta \left ( \text{e}^{\text{i} s} - 
  \frac{ \text{Z}[\psi_{r,s}]}{|\text{Z}[\psi_{r,s}]|} \right ) \; .
\label{holevoDist}\end{equation} 
The corresponding first moment is given by 
\begin{equation}
\langle \text{e}^{\text{i}\beta}\rangle \equiv \langle \text{e}^{\text{i}s} \rangle_H = 
 \sum_{r,s } p_r  q_s  
 \text{e}^{\text{i} \beta [\psi_{r,s}]} \; .
\label{holevoDef} \end{equation} 
An advantage of this expression is that its phase can be considered
as a mean GP, while its modulus is related to a measure
\begin{equation} 
  W = | \langle \text{e}^{\text{i}\beta}\rangle |^{-2} -1 
\label{holevoMeasure} \end{equation} 
for the spread of the GP. 
Eqs.~(\ref{berrydef}) and (\ref{holevoDef}) 
yield different results for the mean GP, which
reflects the choice available in obtaining an average for the GP
from the same distribution.
Ideally, in each run one of the states $|\psi_{r,s} \rangle $ is realized
and leads to a well defined complex value for
$Z[\psi_{r, s}]$.
It is then only a matter of definition how the average of
the complex values over all runs is performed.

{\em Initially pure systems and density matrix decomposition.}---
To concentrate on the effects of the reservoir we 
focus on an initially pure system with 
$\rho_\text{S}(0) = |\psi_\text{S} \rangle \langle \psi_\text{S} | $, so that
\begin{equation} 
  P_\text{Z}(z) = \sum_{r} p_r \delta (z - \text{Z}[\psi_{r}]) ,
\label{berrySimple}\end{equation} 
with $|\psi_{r}(t)\rangle = 
\langle r | U_\text{SR}(t) |r \rangle |\psi_\text{S} \rangle$.
An immediate consequence for a reservoir in a pure state is
that the GP distribution is sharp (i.e.\ a $\delta$ function).
An important example of the reservoir being initially in
a pure state is any energy state, in particular its ground state.

For the reservoir in a mixed state, however,
the non-linear dependence of the dynamic phase functional $D[\psi]$
on $|\psi \rangle$ leads to an ambiguity in Eqs.~(\ref{berrySimple}) and
(\ref{holevoDist}) if the density matrix can be decomposed
into mixtures of two different sets of states \cite{kult03}.
The GP then not only depends on the choice of distribution but
also on the density matrix decomposition.
We argue here that, with respect to the mean GP, 
the introduction of physical constraints can resolve
both ambiguities.
Naturally-occurring reservoirs do not exhibit coherence
between different energy levels, so we assume the density is 
block diagonal in the energy basis (with block sizes determined by
degeneracies);  the thermal reservoir with density operator
$\exp(-\beta H_\text{R})$ is a typical example of such a 
reservoir density matrix. It is then physically reasonable to only admit
decompositions of the density matrix which differ with respect
to the decomposition in degenerate subspaces.
As the dynamic phase functional $D(E)$ is identical for all states 
sharing the same energy eigenvalue $E$, 
the contribution of the respective subspace ${\cal H}_E$ takes the form
\begin{equation} 
\sum_{r \in {\cal H}_E} p_r \delta (z - D(E) 
   \langle \psi_\text{S}| \langle r | U_\text{SR} | r \rangle 
    |\psi_\text{S} \rangle   ) \; .
\end{equation} 
Hence, the corresponding contribution to $\langle z \rangle_\text{Z} $ 
can be written
as $\mbox{Tr}_{\text{S},{\cal H}_E} (U_\text{SR} \rho(0))$, which is
independent of the decomposition. This is not the case 
for the mean GP in the Holevo measure, 
so that the resolution of the decomposition
problem favors the choice of the measure $P_\text{Z}(z)$ for a
GP distribution. We note that for
higher moments the decomposition ambiguity remains for both choices
of GP distribution. However, we will see below that 
for a weakly coupled reservoir the GP distributions coincide
and are independent of the decomposition.

{\em Explicit expression for a weakly coupled reservoir.}---
To derive an explicit expression for GP when the system's evolution
is described by a family of CP maps,
we start with a total hamiltonian of the form
$ H(t) = H_\text{S}(t) + H_\text{R} + H_\text{I}$  
with constant weak coupling $H_\text{I}$ and time-dependent
system hamiltonian $H_\text{S}(t)$,
which generates a unitary evolution
$U_\text{SR}(t) = T\exp (-\text{i} \int_0^t \text{d}t^\prime H(t^\prime ))$
of the total system. 
The reservoir is initially in a mixture of eigenstates $ |r \rangle$
of the time-independent hamiltonian $H_\text{R}$.
In the interaction picture, 
$ U_\text{SR}(t) = U_\text{S}(t) U_\text{R}(t) \tilde{U}(t)$,
with $U_\text{R}(t) = \exp (-\text{i} t H_\text{R})$ and
$U_\text{S}(t) = 
T\exp (-\text{i} \int_0^t \text{d}t^\prime H_\text{S}(t^\prime ))$.
Standard second-order perturbation theory then leads to
$ \tilde{U}(t)= 1 + A + B + O(H_\text{I}^2)$ 
with $A\equiv -\text{i} \int_0^t \text{d}t^\prime 
\tilde{H}_\text{I}(t^\prime )$
as well as 
$B \equiv - \int_0^t \text{d}t^\prime  \int_0^{t^\prime}  
\text{d}t^{\prime \prime}
\tilde{H}_\text{I}(t^\prime ) \tilde{H}_\text{I}(t^{\prime \prime } )$
and the interaction-picture hamiltonian
$\tilde{H}_\text{I} = U_\text{R}^\dagger U_\text{S}^\dagger H_\text{I}
U_\text{S} U_\text{R}$.

Instead of explicitly deriving closed expressions for the
operators $A$ and $B$, we relate them to a
corresponding CP mapping of the reduced density matrix $\rho_\text{S}$,
which is assumed to be of Lindblad type \cite{lindblad77},
\begin{equation}
\dot{\rho}_\text{S} = -\text{i} [H_\text{S} + \Delta H, \rho_\text{S}] 
  - \sum_\alpha ( L_\alpha^\dagger L_\alpha
  \rho_\text{S} + \rho_\text{S} 
  L_\alpha^\dagger L_\alpha - 2  L_\alpha \rho_\text{S}
  L_\alpha^\dagger )
\label{lindblad}\end{equation} 
where $L_\alpha$ are the jump operators. 
The  hermitean operator 
$\Delta H$ describes any energy 
shifts (such as the Lamb shift for atom-light interaction)
associated with S+R interaction.
On the other hand, we can also directly calculate $\dot{\rho_\text{S}}$ 
to second order in $H_\text{I}$ by use of Eq.~(\ref{rdmdef}). An
elementary calculation leads to
\begin{eqnarray} 
  \dot{\rho}_\text{S} &=& -\text{i} [H_\text{S} , \rho_\text{S} ] + 
  \Big \{
  U_\text{S} \langle \dot{B}\rangle_\text{R} 
   U_\text{S}^\dagger  \rho_\text{S}
  \label{lindblad2ndorder} \\ & &
  + U_\text{S} 
   \sum_{r,r^\prime } p_r 
  \langle r|\dot{A} |r^\prime \rangle  
  U_\text{S}^\dagger \rho_\text{S} U_\text{S}
     \langle r^\prime |A^\dagger | r \rangle 
 U_\text{S}^\dagger 
 + \text{H.c.} \Big \}\; .
\nonumber\end{eqnarray} 
Expectation values are denoted by  $\langle \cdots \rangle_q $
for a state $| \psi_q \rangle $; in particular, 
$\langle \cdots \rangle_\text{S} $ refers to
$\langle \psi_\text{S}|  \cdots |\psi_\text{S} \rangle $
and 
$\langle \cdots \rangle_\text{R} $  
to $\mbox{Tr}_\text{R} (\rho_\text{R}(0) \cdots)$.
Comparing Eq.~(\ref{lindblad2ndorder})
with the Lindblad form (\ref{lindblad}) allows us to identify
\begin{equation} 
   U_\text{S}  \langle \dot{B}\rangle_\text{R} U_\text{S}^\dagger =
   -\text{i} \Delta H  - \sum_\alpha  L_\alpha^\dagger L_\alpha 
\end{equation} 
while the operator $A$ is related to the jump terms in a non-trivial way.

To keep the presentation concise we now focus on a
coupling of the form 
$H_\text{I} = - \sum_\mu R_\mu S_\mu$, where $R_\mu$ and $S_\mu$
are operators which act only on ${\cal H}_\text{R}$ and ${\cal H}_\text{S}$,
respectively. Furthermore, we assume that 
\begin{equation} 
  \langle r | R_\mu | r \rangle =0 \; \forall r \; ,
\label{rcond} \end{equation} 
which is the case for energy-transfering S+R interactions, for
instance. An immediate consequence is that $\langle A \rangle_r =0$.
It follows that 
$Z[\psi_r] = Z[\psi_\text{S}]  
(1 + \langle \Delta \text{Z} \rangle_{r,\text{S}} )$ with
\begin{equation} 
   \Delta \text{Z} \equiv  
    \frac{U_\text{S} B }{\langle U_\text{S}\rangle_{\text{S}} }
   - \frac{B - B^\dagger}{2}    
   +\text{i} \int_0^t \text{d}t^\prime
  ( B^\dagger \Delta\tilde{H}_\text{S} + 
  \Delta\tilde{H}_\text{S} B )
   \; ,
\label{zrdef}
\end{equation} 
with
$\tilde{H}_\text{S} \equiv U_\text{S}^\dagger H_\text{S} U_\text{S}$
and 
$\Delta \tilde{H}_\text{S} \equiv 
\tilde{H}_\text{S}-\langle\tilde{H}_\text{S}\rangle_\text{S}$.
Using Eq.~(\ref{zrdef}) one easily finds the following expression
for the moments associated with each of the two different GP distributions
introduced above,
\begin{equation}
  \langle \text{e}^{\text{i} n s} \rangle_\text{H} = 
  \frac{\langle z^n \rangle_\text{Z}}{|\langle z \rangle_\text{Z}|^n}
  = \text{e}^{\text{i} n \beta [\psi_\text{S}] } (1 + \text{i} n \text{Im} 
  \langle \Delta \text{Z} \rangle_{\rho_\text{SR}(0)}  ) \; .
\label{berryResult}\end{equation} 
Result (\ref{berryResult}) has some interesting consequences.
First, $P_\text{Z}$ and $P_\text{H}$ generate exactly the same moments 
so that $ \langle \text{e}^{\text{i} \beta} \rangle =
 \text{e}^{\text{i} \langle \beta \rangle } $ to second order in $H_\text{I}$. Hence
for a weakly coupled reservoir $P_\text{Z}$ and $P_\text{H}$ are equivalent
and, since $|\langle \text{e}^{\text{i} \beta} \rangle|\approx 1$, the GP distribution
is sharp.
Second, as a direct consequence of Eq.~(\ref{rcond}), 
$P_\text{Z}$ and $P_\text{H}$ do not depend on the operator $A$,
so that the GP distribution does not
depend on the jump operators appearing in the master equation.
Third, as a consequence of the ``linearity'' of expressions to lowest
non-trivial order in perturbation theory,
the moments (\ref{berryResult})
are invariant under a change of the reservoir basis. Hence,
they are independent of the 
decomposition of $\rho_\text{R}(0)$.

{\em Explicit Calculations of Berry phase.}---
As illustration we consider the GP distribution 
defined above for some specific 
physical examples. First, we discuss the case of a two-level atom
with ground state $| g \rangle $ and excited state $| e \rangle $
interacting with a thermal radiation reservoir.
The corresponding hamiltonian is given by
$H_\text{S}=
-(\hbar\omega/2) (| e \rangle \langle e | - | g \rangle \langle g| )$,
and the jump operators of Eq.~(\ref{lindblad}) are 
given by $L_1 = \sqrt{\gamma_0 (n+1)} |g \rangle \langle e|$ and
$L_2 = \sqrt{\gamma_0 n} |e \rangle \langle g|$. Here, $\gamma_0$ 
denotes the
spontaneous emission rate and $n$ the thermal mean number
of resonant photons \cite{wallsMilburn}. 
At temperature $T=0$ we have $n=0$ so that this model 
reduces to a description of spontaneous emission in vacuum. 
The operator $B$ introduced above then reads
$B = -\gamma_0 ( | e \rangle \langle e | + n {\bf 1}) $. 
For simplicity we have omitted the 
Lamb shift \cite{lambRemark}. 
For an initial state of the form 
$|\psi_\text{S} \rangle = \cos\frac{\theta}{2} |e \rangle +
  \sin\frac{\theta}{2} |g \rangle$ we find for the (sharp) GP at time
$t= 2\pi/\omega$ the expression 
\begin{equation} 
  \langle \beta \rangle  = \beta[\psi_\text{S}]
   + \pi^2
   \frac{\gamma}{\omega} \sin^2 \theta \; ,
\label{betaReservoir}\end{equation}
where the GP for a closed system is given by 
$\beta[\psi_\text{S}] =  2\pi \sin^2 (\theta /2)$.
An interesting feature of this result is that the mean GP
does not depend on the temperature, even though for $T>0$ the
radiation reservoir is in a mixed state. 
This is a consequence of the $n$-dependence of $B$ being proportional
to the identity: the 
effect of thermal fluctuations, which induce incoherent absorption 
and emission of thermal photons at equal rates, do 
exactly cancel each other. The asymmetric effect of spontaneous
emission, however, leads to a change in the GP. 

Alternatively, one can calculate the GP by solving 
Eq.~(\ref{lindblad}) in terms of Kraus operators $K_i$, 
without explicit reference to
the reservoir, which provides an exact calculation of the
GP distribution. This simply amounts to seeking for operators $K_i$ and
probability weights $p_i$ for which
$\sum_i p_i K_i \rho_\text{S}(0) K_i^\dagger$ is a solution 
of the master equation \cite{krausRemark}.
For the present case, the Kraus operators are given by

\begin{eqnarray} 
 K_0 &=& \text{e}^{-\text{i} \frac{\omega}{2}t} |g \rangle \langle g| +
      \text{e}^{\text{i} \frac{\omega}{2} t -\gamma_n t} |e \rangle \langle e|
 \\
 K_1 &=& \sqrt{1-\text{e}^{-2\gamma_n t}} | g \rangle \langle e |
  \\
 K_2 &=& \text{e}^{-\text{i} \frac{\omega}{2}t-\gamma_n t} |g \rangle \langle g| +
      \text{e}^{\text{i} \frac{\omega}{2} t } |e \rangle \langle e|
  \\
 K_3 &=& \sqrt{1-\text{e}^{-2\gamma_n t}} | e \rangle \langle g |
\end{eqnarray} 
with $\gamma_n \equiv (2n+1)\gamma_0$ and weights $p_0=p_1= (n+1)/(2n+1)$
as well as $p_2=p_3= n/(2n+1)$.
The operators $K_1$ and $K_3$ are related to the jump operators
and, because of $K_1(0)=K_3(0)=0$, do not contribute to the GP. 
Substituting $K_0$ and $K_2$ into Eqs.~(\ref{berrydef}) and
(\ref{holevoDef})
we calculate the GP distribution at time
$t= 2\pi/\omega$ to be
\begin{eqnarray}  
  P_\text{Z}(z) &=& 
     p_0 \delta (z-f_-) + p_2 \delta (z-f_+) 
  \nonumber \\
  P_\text{H}(s) &=& 
     p_0 \delta \left (\text{e}^{\text{i} s} -\frac{f_-}{|f_-|}\right) 
   + p_2 \delta \left (\text{e}^{\text{i} s} -\frac{f_+}{|f_+|} \right) 
 \\
 f_\pm &\equiv& - \text{e}^{-\pi \frac{\gamma_n}{\omega}} 
  \langle \text{e}^{\mp \pi \frac{\gamma_n}{\omega} \sigma_z} \rangle_\text{S}
  \;
  \langle \text{e}^{\mp 2\pi \frac{\gamma_n}{\omega} \sigma_z} 
  \rangle_\text{S}^{\pm \text{i} \frac{\omega}{2\gamma_n}  } \; .
\nonumber\end{eqnarray} 
For zero temperature we have $n=0$ and therefore $p_2=0$. Both
expressions then predict a sharp GP 
\begin{equation}
  \langle \beta \rangle = \pi + \frac{\omega}{2\gamma_0}
    \ln \langle \psi_\text{S} | \text{e}^{-2\pi \gamma_0 \sigma_z /\omega} |\psi_\text{S} \rangle \; .
\end{equation}
This result agrees with the expression found in Ref.~\cite{fuentes03} and,
to first order in $\gamma_0$, also with the result (\ref{betaReservoir})
based on the weakly coupled reservoir. Any difference between
$\exp \text{i} \langle \beta \rangle $ and 
$\langle \exp \text{i} \beta \rangle $
is of second order in $\gamma_0$. 
Also for finite temperatures, the two exact results still agree with 
the weak coupling  result (\ref{betaReservoir}) to first order in 
$\gamma_0$. Hence, any dependence on the temperature through $n$ is of 
higher order in $\gamma_0$.

Another illustrative case is that of phase damping which can be 
described by a jump operator of the form 
$L_1 = \sqrt{\alpha}( |e \rangle \langle e| -|g \rangle \langle g|)$
(and therefore $B \propto {\bf 1}$),
where $\alpha$ denotes the phase damping rate. 
This jump operator
can be derived from a coupling to a non-resonant
reservoir of harmonic oscillators with
effective interaction hamiltonian 
$H_\text{I} = S_0 R_0 = \sigma_z \sum_i g_i a_i^\dagger a_i$, 
where $a_i$ is the annihilation
operator of the $i$th oscillator and $g_i$ the corresponding effective
coupling parameter. In thermal equilibrium we have
$\langle R \rangle = \sum_i g_i \langle a_i^\dagger a_i \rangle \neq 0$
so that condition (\ref{rcond}) is violated. Consequently the (trivial)
result predicted by Eq.~(\ref{berryResult}) is spurious.

We can again compare this to an exact calculation 
based on the Kraus operators 
\begin{eqnarray} 
  K_0 &=& \frac{1}{r} \text{e}^{-\text{i} \frac{\omega}{2}t-\alpha t} 
  |g \rangle \langle g| +
  r \text{e}^{\text{i} \frac{\omega}{2} t} |e \rangle \langle e|
  \\
 K_1 &=& r \text{e}^{-\text{i} \frac{\omega}{2}t} |g \rangle \langle g| +
      \frac{1}{r} 
     \text{e}^{\text{i} \frac{\omega}{2} t -\alpha t } |e \rangle \langle e|
\end{eqnarray} 
with $r \equiv (1+\sqrt{1-\exp(-2\alpha t)})^{1/2}$ and weights
$p_0=p_1=1/2$. For brevity we only discuss the first moments 
which are given by
\vspace*{-4mm}
\begin{eqnarray} 
  \text{e}^{\text{i} \langle \beta \rangle } &\approx& 
   \text{e}^{\text{i} \beta[\psi_\text{S}] }
  \left ( 1+ \frac{2\text{i} \pi^2 \alpha}{3\omega} \cos\theta \sin^2\!\theta
  \right ) 
  \\ 
  \langle \text{e}^{\text{i}\beta } \rangle &\approx&
\text{e}^{\text{i} \beta[\psi_\text{S}]}
  \left (\! 1\! + \! \frac{2\pi^2 \alpha}{\omega} \sin^2\!\theta
     (\text{i}  \cos\theta \!- \! \frac{4}{9} \sin^2\!\theta  )
  \right ) 
\end{eqnarray} 
for $t=2\pi/\omega$. In contrast to Eq.~(\ref{berryResult}),
these moments include non-trivial
corrections and differ from each other
as well as from the result of Ref.~\cite{fuentes03}.
By Eq.~(\ref{holevoMeasure}) they indicate a GP spread of
$W \approx 16\pi^2 \sin^4\theta \alpha/(9\omega) $ for phase damping.

In summary we have established a theory for GP distributions based
on operational considerations that reduces to the result
of Samuel and Bhandari \cite{samuel88} for no-jump non-unitary evolution
and employs a Kraus operator analysis. We address and resolve
ambiguities concerning decomposition of the density matrix
and GP by incorporating reasonable assumptions 
about the reservoir and solve specifically for spontaneous
emission and phase damping of a two-level atom.
The resultant GP distributions are derived and the ambiguities over
phase mean and spread are consequences of choices of how to define
a phase distribution. The underlying method  for defining the
GP distribution is now clear, but of course choices of the
distributions are ultimately determined by experimental considerations.
Future work will consider the GP distribution for non-abelian GP.
\\[0mm]
{\bf Acknowledgments}
We thank J. Watrous for helpful discussions. 
BCS acknowledges valuable discussions with J.~D.~Cresser
and I.~Kamleitner.
This work was supported
by Alberta's informatics Circle of Research Excellence (iCORE).


\begin{thebibliography}{99}
\bibitem{berry84} M.V. Berry, Proc.~Roy.~Soc.~(Lond.) {\bf 392},
        45 (1984).
\bibitem{pancharatnam56} S. Pancharatnam, 
    Proc.~Ind.~Acad.~Sci.~Sect.~A~{\bf 44}, 247 (1956).
\bibitem{anandan88b} J. Anandan,  
        Phys.~Lett.~A {\bf 133}, 171 (1988).
\bibitem{aharonov87} Y. Aharonov and J. Anandan, 
    Phys. Rev. Lett. {\bf 58}, 1593 (1987).
\bibitem{samuel88} 
  J. Samuel and R. Bhandari, Phys.~Rev.~Lett.~{\bf 60}, 2339 (1988).
\bibitem{uhlmann86} A. Uhlmann, Rep. Math. Phys, {\bf 24}, 229 (1986).
\bibitem{pati95a} A.K.~Pati, J.~Phys.A {\bf 28}, 2087 (1995).
\bibitem{pati95b} A.K.~Pati, Phys.~Rev.~A {\bf 52}, 2576 (1995).
\bibitem{adiqc}  E. Farhi {\em et al.},
   quant-ph/0001106;
   A.M. Childs, E. Farhi, and J. Preskill,
   Phys.~Rev.~A {\bf 65}, 012322 (2002).
\bibitem{pachos99} J. Pachos, P. Zanardi, and M. Rasetti,
   Phys.~Rev.~A {\bf 61}, 10305 (1999).
\bibitem{gamliel89} D. Gamliel and J.H. Freed, Phys.~Rev.~A {\bf 39},
  3238 (1989).
\bibitem{ellinas89} D.~Ellinas, S.M.~Barnett, and M.A.~Dupertuis,
   Phys.~Rev.~A {\bf 39}, 3228 (1989).
\bibitem{sjoquist00} 
  E. Sj\"oqvist {\em et al.},
  Phys.~Rev.~Lett.~{\bf 85}, 2845 (2000).
\bibitem{fonseca02} K.M. Fonseca Romero, A.C. Aguiar Pinto and M. T. Thomaz,
  Physica A {\bf 307}, 142 (2002).
\bibitem{ericsson03} M. Ericsson {\em et al.}, 
  Phys.~Rev.~A {\bf 67}, 020101(R) (2003).
\bibitem{pati03} A.~K.~Pati, Int.~J.~of 
  Quantum Information, {\bf 1}, 135 (2003).
\bibitem{peixoto03} J.G. Peixoto de Faria, A.F.R. Toledo Piza, and
  M.C. Nemes, Europhys. Lett. {\bf 62}, 782 (2003).
\bibitem{singh03} K.~Singh {\em et al.},
   Phys.~Rev.~A {\bf 67}, 32106 (2003).
\bibitem{fu04} L.-B.~Fu and J.-L.~Chen,
   J.~Phys.~A: Math.~Gen.~{\bf 37}, 3699 (2004); 
   E. Sj\"oqvist, quant-ph/0404174 (2004); 
   L.-B.~Fu and J.-L.~Chen, quant-ph/0407076 (2004).
\bibitem{whitney04} R.S.~Whitney {\em et al.},
     cond-mat/0401376 (2004).
\bibitem{fuentes03} A. Carollo {\em et al},
  Phys.~Rev.~Lett.~{\bf 90}, 160402 (2003).
\bibitem{fuentes03b}  I. Fuentes-Guridi, F. Girelli, and E. Livine,
    quant-ph/0311164. 
\bibitem{kult03} D. Kult and E. Sj\"oqvist, quant-ph/0312094 (2003).
\bibitem{webb99} C.L.~Webb {\em et al.},
   Phys.~Rev.~A {\bf 60}, R1783 (1999).
\bibitem{sanders01}  B.C.~Sanders {\em et al.},
   Phys.~Rev.~Lett.~{\bf 86}, 369 (2001).
\bibitem{betaMeanRemark} 
   $\exp \text{i} \langle \beta \rangle $ is only
  a notation representing a pure phase factor. 
  Generalizations to higher order moments 
  $\langle z^n\rangle_\text{Z}$ are straightforward but not 
  necessary for this analysis.
\bibitem{holevo84} A.S.~Holevo, Springer Lecture Notes Math.~{\bf 1055},
  153 (1984).
\bibitem{lindblad77} G. Lindblad, Commun.~Math.~Phys.~{\bf 48}, 119 (1976).
\bibitem{wallsMilburn} D. F. Walls and G. J. Milburn, {\em Quantum Optics},
  Springer, Berlin (1995).
\bibitem{lambRemark} The Lamb shift contributes to the GP for
   $t\neq 2\pi/\omega$ only.
\bibitem{krausRemark} $K_i$ are different from
  $\langle r | U_\text{SR} | r \rangle $  introduced earlier.
\end{thebibliography}
\end{document}